\documentclass[twocolumn, amsmath, amssymb,floatfix,notitlepage,nobibnotes,reprint,prx]{revtex4-2}

\usepackage{booktabs}
\usepackage{romannum}

\usepackage{graphicx}

\newcommand{\sht}[1]{\textcolor{blue}{#1}}  
\usepackage{xcolor}
\usepackage{bm}
 \usepackage{textcomp}
\usepackage[colorlinks=true, citecolor=blue, linkcolor=blue, urlcolor=blue]{hyperref}

\usepackage{listings}
\lstdefinelanguage{PythonCustom}{
  language=Python,       
  morekeywords=[1]{import, from, for, in, range},
  morekeywords=[2]{os, collections, cqlib, cqlib_experiments, TianYanPlatform,setup_circuit_with_depth,defaultdict,transpile,run_task,supremacy_result,rcs,job,download_data,show,show_opt_parameters},
  basicstyle=\ttfamily\tiny,               
  backgroundcolor=\color{gray!15},         
  breaklines=true,                         
  tabsize=4,                               
  frame=none,                              
  columns=flexible,                        
  commentstyle=\itshape\color{teal},       
  keywordstyle=[1]\color{green!50!black},  
  keywordstyle=[2]\color{blue},    
  stringstyle=\color{red!80!black},        
  identifierstyle=\color{black},           
}

\begin{document}

\title{{\textit{\textbf{}{Tianyan}}: Cloud services with quantum advantage}}
\author{$Tianyan~Quantum~Group$\\ China Telecom Quantum Information Technology Group Co., Ltd., Hefei 230031, China}

\begin{abstract}
$Tianyan$ Quantum Cloud Platform offers cloud services demonstrating quantum advantage capabilities with a $Zuchongzhi\ 3.0$-like superconducting quantum processor. This cloud-accessible superconducting quantum prototype, named $Tianyan$-287, features 105 qubits and achieves high operational fidelities, with single-qubit gates, two-qubit gates, and readout fidelity at 99.90\%, 99.56\%, 98.7\%, respectively. For a specific benchmark task involving random circuit sampling on a 74-qubit system over 24 cycles, the platform completes one million samples in just 18.4 minutes. In contrast, state-of-the-art classical supercomputers would require approximately 16,000 years to complete the equivalent calculation. To facilitate this, the platform provides access via Cqlib, an open-source SDK designed for working with quantum systems at the level of extended quantum circuits, operators, and primitives. The cloud service aims to democratize access to high-performance quantum hardware, enabling the community to validate and explore practical quantum advantages.

\setlength{\parskip}{\baselineskip} 
$Tianyan$  Quantum Cloud Platform: \sht{\url{https://qc.zdxlz.com}}

\end{abstract}

\date{\today}

\maketitle

\section{Introduction}

Quantum computing has emerged as a fundamental shift in information processing, harnessing the unique properties of quantum mechanics to address computational challenges previously considered intractable for classical systems. We are currently operating within the Noisy Intermediate-Scale Quantum (NISQ) period \cite{greenwade93, preskill2018}, characterized by remarkable progress alongside persistent technical constraints. The emergence of cloud-accessible quantum systems has been instrumental in advancing the field, with multiple organizations now providing remote access to quantum processors. This accessibility has catalyzed unprecedented growth in quantum research, with experimental implementations and theoretical investigations expanding at an accelerating pace across both academic and industrial domains. 

Within this ecosystem, $Tianyan$ Quantum Cloud Platform has firmly established itself as a key contributor through its sustained development of  quantum hardware and software. By integrating the $Tianyan$-176 superconducting quantum system with High-Performance Computing (HPC) cluster, $Tianyan$ achieved hybrid quantum-classical computing in 2023. In 2024, the platform delivered a 504 qubits superconducting quantum system, marking the first validation of a 504-qubit-scale system. Furthermore, $Tianyan$ released an open-source quantum computing programming framework - Cqlib, establishing a full-stack domestic quantum computing software system. These technological advancements have enabled increasingly complex quantum experiments and algorithm implementations across various domains including quantum simulation, optimization, and machine learning.

Quantum computational advantage is a pivotal milestone in the field of quantum computing. In 2019, Google achieved what is now widely recognized as "Quantum advantage" through a random circuit sampling (RCS) task \cite{arute2019quantum}. This marked the first time a quantum system demonstrated the ability to outperform classical supercomputers on specific tasks. In the subsequent years, the technology has entered a rapid development phase, witnessing continuous improvements in both qubit scale and circuit depth \cite{wu2021strong, zhu2022quantum, morvan2024phase, zhong2020quantum, deng2023gaussian, zhong2021phase, madsen2022quantum, google2025, gao2025establishing}. As of now, the largest-scale quantum advantage experiment reported was conducted on the Willow chip, implementing a quantum circuit with 103 qubits and a depth of 40 \cite{google2025}, surpassing the previous record held by $Zuchongzhi$\ 3.0 \cite{gao2025establishing}, which utilized 83 qubits and a depth of 32.

As bottlenecks in quantum hardware are gradually overcome, the core focus of the entire field has shifted from "how to compute faster" to "how to compute usefully"—exploring which practical scientific challenges quantum computing can truly solve. The quantum echo experiment conducted by the Google team has once again marked the beginning of this phase, owning to their quantum hardware system capable of achieving quantum advantage \cite{google2025}. Therefore, possessing such computational advantage power has become a prerequisite for exploring the practical utility of quantum computing. Cloud-access to a superconducting quantum computer with quantum advantage capabilities, can effectively enable a broader range of researchers and engineers participate in the journey toward practical quantum computing. This approach will accelerate the arrival of the era of quantum computing utility. 

In this work, we introduce quantum advantage cloud services via $Tianyan$ platform, leveraging a high-performance superconducting quantum computing system. This system features a $Zuchongzhi\ 3.0$-like processor, integrating 105 physical qubits with exceptionally high-fidelity quantum control. The single-qubit gates, two-qubit gates, and readout fidelities are 99.90\%, 99.56\%, and 98.7\%, respectively. Leveraging this system, we demonstrated 74-qubit RCS experiments with 24 cycles in 18.4 minutes of one million samples. As a contrast, the state-of-the-art classical supercomputer would require an estimated 1.6$\times 10^4$ years to replicate the same sampling endeavor. Through our open-source Cqlib SDK, we democratize access to high quantum computational power, enabling the community to test and explore practical applications.

\begin{figure*}
\centering
\includegraphics[width=0.95\linewidth]{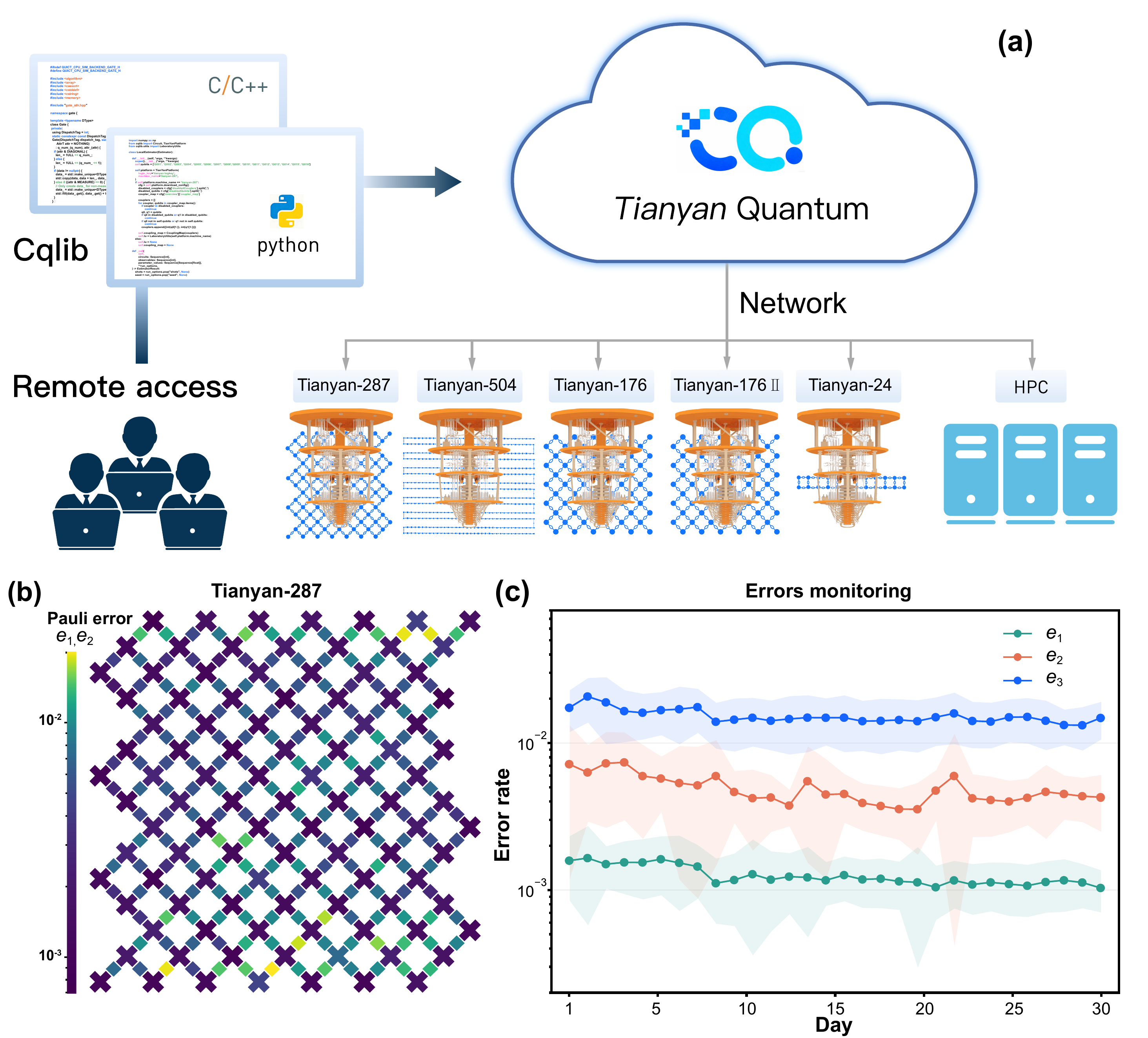}
\caption{\label{fig:1} $Tianyan$ quantum cloud services with high-performance quantum computing systems. a) The illustration of the  $Tianyan$ cloud services, providing access to superconducting quantum computing systems include $Tianyan$-287,  $Tianyan$-176,  $Tianyan$-504 and  $Tianyan$-24. All the systems are cloud-accessible via Cqlib SDK.
b) and c) show the key indicators and its daily fluctuation of the $Tianyan$-287 system, which equipped with a $Zuchongzhi\ 3.0$-like superconducting quantum processor.
b) Heat map showing single- and two-qubit Pauli errors $e_1$ (crosses) and $e_2$ (bars) positioned in the layout of the processor. Values are shown for all qubits operating simultaneously. c) The daily collection of the key indicators for 30 consecutive days, including  single- and two-qubit Pauli errors $e_1$ and $e_2$, and readout error $e_3$, with shaded regions corresponding to the respective uncertainty intervals. }
\end{figure*}

\section{Tianyan high-performance quantum cloud services}

$Tianyan$ Quantum Cloud Platform is among the leading providers of quantum computing resources, offering access to cutting-edge quantum hardware built with advanced technology. $Tianyan$ operates the world's most advanced fleet of utility-scale quantum systems, As is shown in the diagram in  Fig.~\ref{fig:1} a), currently with five systems in operation and more under development. These systems are highly reliable, maintaining stable up-time and remarkable small gate errors. In addition to hardware systems, $Tianyan$ provides comprehensive software resources, via the open-source Cqlib SDK, that support learning, experimentation, and collaboration in quantum computing. 

In 2023, $Tianyan$ introduced two superconducting quantum systems, $Tianyan$-176 and $Tianyan$-176 \Romannum{2}, featuring the previous high-performance $Zuchongzhi$ 2.0-like processor. By integrating the $Tianyan$-176 systems  with high-performance computing (HPC) clusters, $Tianyan$ enables quantum-classical hybrid computing. Then in 2024, $Tianyan$ launched the  $Tianyan$-504 superconducting quantum system with 504 physical qubits, achieving a previous milestone of the largest commercially available quantum computer. This demonstrated an engineering exploration on system scaling limitation. 

Here we release the latest system launched on the $Tianyan$ Cloud Platform --- $Tianyan$-287 system, equipped with a $Zuchongzhi\ 3.0$-like superconducting quantum processor. This prototype integrates 105 qubits and 182 couplers in a square grid lattice. The qubits have a mean operating T$_1$ of 44.4 $\mu$s and T$_2^{CPMG}$ of 41.1 $\mu$s. Increasing coherence contributes to all the fidelity of qubit gate operations which are displayed in Fig.~\ref{fig:1} b). Heat map showing single- and two-qubit Pauli errors $e_1$ (crosses) and $e_2$ (bars) positioned in the layout of the processor. For all qubits operating simultaneously, the average value of $e_1$ and $e_2$ are 1\textperthousand ~and 4.4\textperthousand, respectively. The average readout error $e_3$ is 1.3\%.

To assess the system's long-term reliability, we monitored key error parameters from August 29 to September 28, 2025 --- a 30-day period aligning with the 1–30 day dimension in Fig.~\ref{fig:1} c). The evaluation focused on three core errors: $e_1$, $e_2$ and $e_3$. System stability was determined from daily average trends and fluctuation ranges of these indicators. The observed long-term stability, including uncertainty intervals depicted as shaded regions, verifies consistent system performance throughout the monitoring period.

\section{RCS experiments}

We showcase the performance of our quantum systems through an experiment known as random circuit sampling. To further widen the performance gap between quantum and classical computation, random circuits must be both random and structurally optimized. In our design, two-qubit gate sequences are arranged to induce quantum behaviors that are particularly hard for classical computers to simulate, following two principles: adapting the layout to the chip architecture and maximizing classical simulation cost \cite{17huang2024design}. The iSWAP-like two-qubit gates are applied according to predefined patterns (labeled A, B, C, and D) and executed in the sequence ABCD–CDAB within each cycle. In each cycle, single-qubit gates are randomly chosen from the set ${\sqrt{X},\sqrt{Y},\sqrt{W}}$.

\begin{figure*}[htbp]
\centering
\includegraphics[width=1\linewidth]{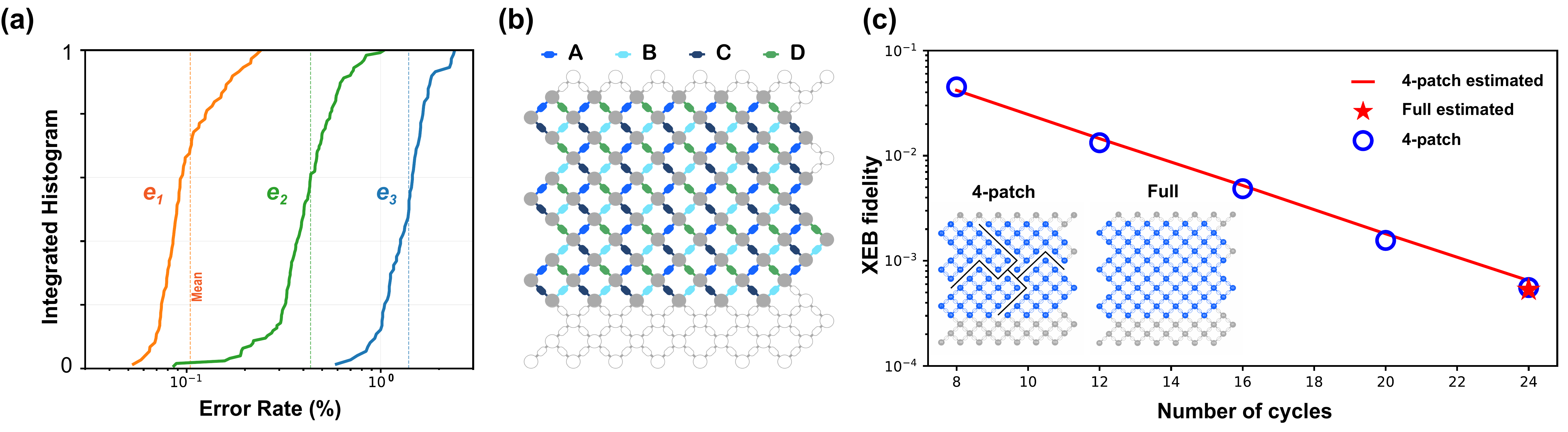}
\caption{\label{fig:Hard}Regional system errors and fidelity of random circuit sampling experiment for 74 qubits. a) Integrated histogram
(cumulative distribution function, CDF) of Pauli errors (orange, green) and readout errors (blue), measured on the 74 qubits region shown in b), when operating all qubits simultaneously. The vertical dotted lines denote the average (mean) values, with the Pauli error for single-qubit gates ($e_1$), two-qubit gates ($e_2$), and readout error measuring 1.0‰, 4.4‰, and 1.3\%, respectively. 
b) The pattern diagram of the 74-qubit random circuit sampling experiment. The iSWAP-like gates are selected from four distinct configurations (A-D) and arranged in the sequence ABCDCDBA. The grey circles denote functional qubits, while blue, cyan, indigo, and green rounded bars represent the iSWAP-like gates associated with the four patterns A, B, C, and D respectively. Additionally, the remaining empty circles and rounded bars are the discarded qubits and couplers. c) Experimental and estimated fidelity of of random circuit sampling experiment for 74 qubits. 
The blue circles and red line represent the experimental and estimated values, respectively, of the 74-qubit 4-patch circuit. The red five-pointed star signifies the estimated value of the 74-qubit full circuit, where 91.255 million bitstrings are sampled.
The inserted topological diagram depicts the specific configuration of 74 qubits. 
}
\end{figure*}

Verifying the fidelity of complete random quantum circuits is challenging, as their ideal probability distributions is hard to be efficiently simulated classically. To address this, we use \textit{patch circuits}, constructed by selectively removing two-qubit gates between regions, to validate large-scale circuits. We implemented two types: a four-patch version and a full 74-qubit circuit, each with 8–24 layers. Linear cross-entropy benchmarking (XEB) fidelities were measured for both. As shown in Fig.~\ref{fig:1}(d), random circuit sampling was successfully demonstrated on the 74-qubit circuit with 12–24 layers. For the largest circuit—the full 74-qubit, 24-cycle configuration—approximately $9.13 \times 10^7$ bitstrings were collected. The 4-patch circuit exhibited experimental and estimated fidelities of 0.056\% and 0.064\%, respectively, confirming that the discrete error model reliably predicts fidelity even at large scale. The fidelity of the full circuit is thus estimated at 0.054\%.


The tensor network algorithm \cite{markov2008simulating,guo2019general,20villalonga2019,villalonga2020,huang2020classical,pan2022simulation,guo2021verifying,pan2022solving,fu2024surpassing,zhao2025leapfrogging,guan2024single} currently represents the state of the art in classical simulation of random quantum circuits. We applied this method \cite{zhao2025leapfrogging,guan2024single} to estimate the classical cost of the most demanding circuit (74 qubits, 24 cycles) under two memory-constrained scenarios. In one of these scenarios, the memory limit is set to 9.2 PB, corresponding to the capacity of Frontier, the world’s most powerful supercomputer. For this circuit, the estimated number of floating-point operations (FLOPs) required to generate 1 million independent bit strings at 0.054\% fidelity is $2.1 \times 10^{28}$. Using Frontier’s theoretical peak performance of $1.685 \times 10^{18}$ single-precision FLOPs per second, assuming 20\% floating-point efficiency, accounting for low target fidelity, and considering that each single-precision complex FLOP requires 8 machine FLOPs, simulating the circuit is estimated to take $1.6 \times 10^{4}$ years.

Since the tensor network algorithm is heavily constrained by memory, we considered a nearly unlimited memory scenario (serving as a lower bound for sampling costs, though not practically feasible), setting the memory limit to over 762.2 PB (including Frontier’s total memory and storage). Under this scenario, generating 1 million bitstrings at 0.054\% fidelity would still require $1.8 \times 10^{25}$ FLOPs, corresponding to an estimated classical simulation time of 19 years. The simulation times in both scenarios far exceed practical limits, highlighting the robustness of the quantum advantage achieved.

\section{Task workflow on \textit{\textbf{}{Tianyan}} via Cqlib}

We provide Cqlib --- an open-source software development kit (SDK) for executing algorithms on quantum systems \cite{cqlib_guide}. Through the SDK, users gain easy access to high-performance quantum systems on the $Tianyan$ cloud platform. It supports advanced quantum tasks including extended circuits, operators, and primitives, providing an end-to-end toolkit for scientific experiments and software developments.

\begin{figure*}[htbp]
\centering
\includegraphics[width=1.0\linewidth]{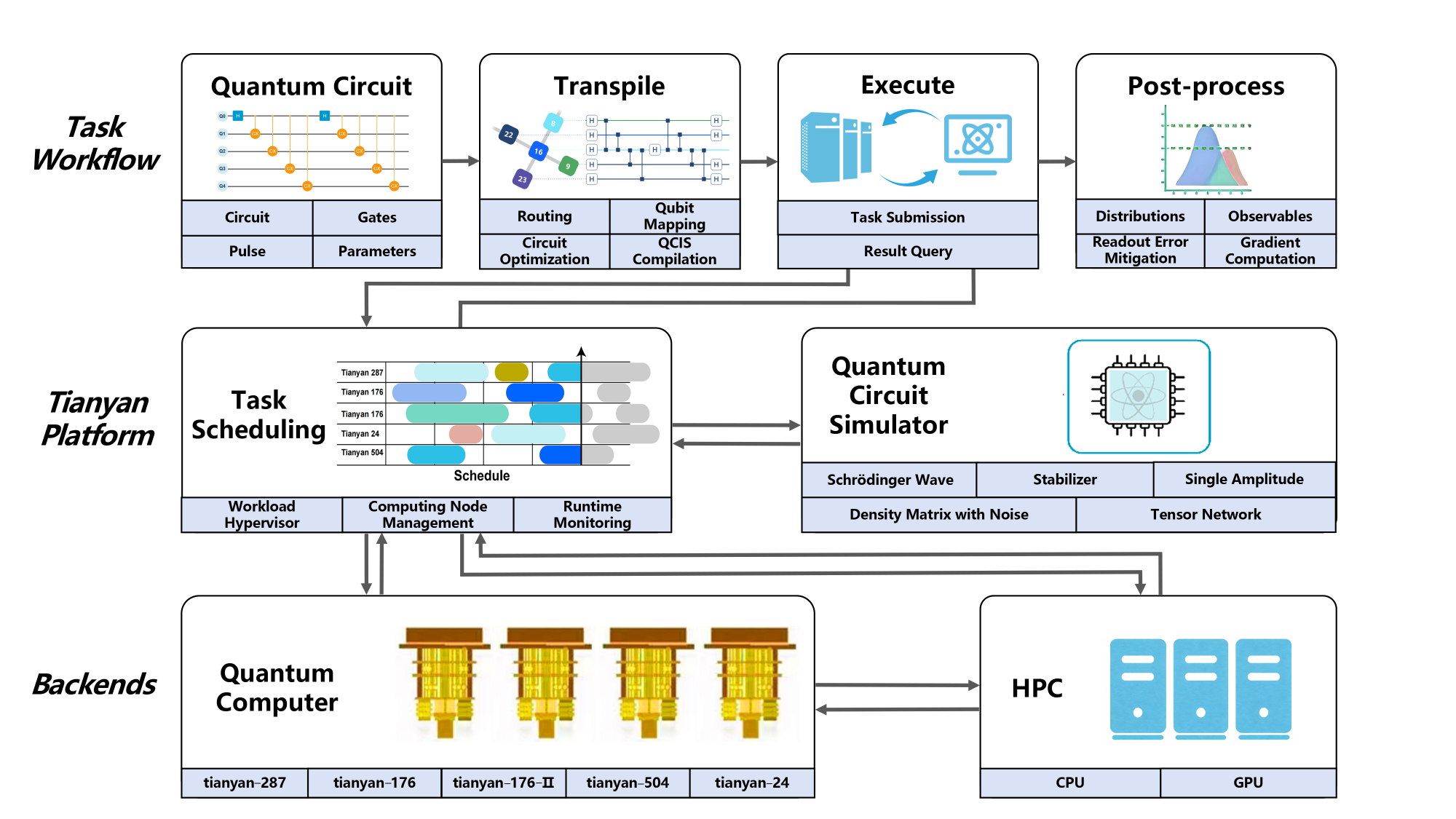}
\caption{\label{fig:3} Cqlib task workflow of executing algorithms on $Tianyan$ platform. The workflow outlines four main steps, including circuit construction, transpilation, execution, and post-processing. Through task scheduling and quantum circuit simulation services on $Tianyan$, hybrid tasks can be executed in an iterative manner across quantum systems and HPC backends. Local low-latency interaction between quantum and HPC systems is achieved through compact deployment. }
\end{figure*}

Cqlib outlines a four-step process for executing tasks on $Tianyan$, compatible with its software architecture shown in Fig. \ref{fig:3}. The process begins by translating classical problems into quantum circuits constructed through high-level abstractions. These circuits then undergo a transpilation phase, where qubit mapping and structural optimizations adapt them to the constraints of the target hardware. Following transpilation, tasks are submitted to $Tianyan$ platform, whose scheduling system manages task queuing and distributes workloads across heterogeneous resources, including quantum circuit simulators, HPC servers, and quantum hardware systems. During task execution, HPC servers act as the classical collaborator, coordinating circuit dispatch, sampling collection, and associated data-processing routines. Finally, the obtained results are post-processed to extract the final solutions to the original problem. 

Leverages the Cqlib four-step workflow, $Tianyan$ platform enables efficient execution of large-scale tasks with circuits that utilize more than 100 qubits and incorporate over 2000 gates. To demonstrate the process, we illustrate the large-scale RCS experiments step by step as an example.

Step 1: Circuit Construction. The process begin by generating random quantum circuits using Cqlib-experiments toolkit. By specifying the number of qubits, the connectivity pattern, and circuit depths, multiple independent circuit instances are generated with evaluated sampling complexity.

\begin{center}
\begin{minipage}{1.0\linewidth}
\begin{lstlisting}[language=PythonCustom]
# Step 1
from cqlib_experiments.rcs.circuit import setup_circuit_with_depth

# The list of logic qubits.
qubits = [1, 2, 3, 4, 5, 6, 7, 8, 9, 10]
# Defines how to connect qubits in two-qubit gates for each layer. A, B, C, D are 
# different connection patterns.
pattern_qubits = {
    'A': [(1, 4), (2, 5), (3, 6)],
    'B': [(5, 8), (6, 9), (7, 10)],
    'C': [(4, 8), (5, 9), (6, 10)],
    'D': [(1, 5), (2, 6), (3, 7)],
}
# The different circuit complexities we want to test.
depths = [12, 16, 20, 24]
# How many random circuits to create for each depth.
repeat = 10

# Create a dictionary of all the circuits to run. For each depth, create a list
# of 'repeat' number of random circuits.
circuits = {
    depth: [
        setup_circuit_with_depth(qubits, pattern_qubits, depth=depth)
        for _ in range(repeat)
    ] for depth in depths
}
\end{lstlisting}
\end{minipage}
\end{center}

Step 2: Transpilation. The generated logical circuits are translated into hardware-compatible instructions according with the topology and gates parameters of the $tianyan$-287 system. This step performs circuit-to-circuit rewriting, producing circuits instructions that can directly execute on the hardware.

Step 3: Execution. The transpiled circuits are then submitted to the $tianyan$-287 system via $Tianyan$ cloud login-key. Each circuit is sampled extensively on the target quantum system. The obtained output distributions under-go an automatic optimization module on the locally-connected HPC to refine the iSWAP-like gate parameters.

\begin{center}
\begin{minipage}{1.0\linewidth}
\begin{lstlisting}[language=PythonCustom]
# Step 2
import os
from collections import defaultdict
from cqlib import TianYanPlatform
from cqlib.mapping import transpile

# Initialize the TianYanPlatform object.
pf = TianYanPlatform(login_key=os.getenv('APIKEY'))
# Download the hardware configuration for tianyan-287
config = pf.download_config(machine='tianyan-287')
# Use a defaultdict to easily store lists of compiled circuits for each depth.
transpiled_circuits = defaultdict(list)

# Iterate through the previously generated circuits.
for depth, circuit_list in circuits.items():
    for circuit in circuit_list:
        # The core step: Transpile the circuit.This process maps the logical circuit to the 
        # physical hardware,
        transpiled_circuit = transpile(circuit, config)
        transpiled_circuits[depth].append(transpiled_circuit)
\end{lstlisting}
\end{minipage}
\end{center}

\begin{center}
\begin{minipage}{1.0\linewidth}
\begin{lstlisting}[language=PythonCustom]
# Step 3
from cqlib_experiments.rcs.job import run_task, supremacy_result

# Define the number of samples (shots) to collect for each circuit. More shots provide
# a more accurate statistical picture of the output distribution.
samples = 30000

# Submit the batch of transpiled circuits to the quantum backend for execution.
# - `auto_opt=True`: After sampling, automatically perform a parameter optimization
#   for the 'fsim' gate model to improve the fidelity of the simulation.
# The function returns a 'task_id' to track the job.
task_id = run_task(config, transpiled_circuits, shots=samples, auto_opt=True)

# Wait for the execution to complete and retrieve the final results. This function 
# will block until the job is finished and the data is processed.
result = supremacy_result(task_id)
\end{lstlisting}
\end{minipage}
\end{center}

Step 4: Post-Processing. Finally, the sampling data-set and the optimized circuit/gate parameters are loaded from the cloud storage. Post-Processing procedures are applied to evaluate final fidelity using XEB model and results visualization.

\begin{center}
\begin{minipage}{1.0\linewidth}
\begin{lstlisting}[language=PythonCustom]
# Step 4
from cqlib_experiments.rcs.job import download_data
from cqlib_experiments.rcs.show import show_opt_parameters

# Download the complete results package from the backend for the given task. 
# This package typically includes the raw sampling results, circuit metadata, 
# and the optimization logs. The data is saved as a zip file for archival and analysis.
download_data(task_id, file_name=f'{task_id}.zip')

# Generate a visualization of the parameter optimization process.
fig = show_opt_parameters(config, task_id)
fig.show()
\end{lstlisting}
\end{minipage}
\end{center}

\section{Conclusion and Outlook}
We have presented $Tianyan$, a quantum cloud platform that provide high-performance quantum computing services and a comprehensive open-source SDK for executing algorithms on quantum systems. Five sets of high-performance quantum computing systems equipped with $Zuchongzhi$- and $Xiaohong$- series superconducting quantum processors are accessible via the cloud. We outlined a four-step quantum program development and execution framework in Cqlib, enabling easy exploration of quantum application development. Leveraging the $Tianyan$-287 cloud service and Cqlib toolkit, we demonstrated a large-scale RCS experiment with 74 qubits and 24 cycles, achieving quantum advantage beyond existing top supercomputers. This service marks the first commercial cloud-access to leading quantum computing power. 

However, RCS tasks based on iSWAP-like gates have limited practical application. But the realization of large-scale quantum algorithms via cloud service showcases the possiblity of transform the experimental technologies on high-precision quantum operation into commercial service capabilities. Technologies can be adapted to other gate-types or algorithms, such as the auto-calibration routine used to maintain the high performance state, the HPC-engaged low-latency iterative optimization, and the phase compensation technologies in large-scale circuit execution. 

We invite quantum developers, researchers, and enterprises to explore and leverage our quantum computational advantage cloud services. Our offering provides a pragmatic, open pathway to access and harness quantum superiority, advancing the exploration of near-term NISQ applications and future large-scale fault-tolerant quantum computing.


\section {Acknowledgments} The authors acknowledge support from Superconducting Quantum Computing Group of Excellence in Quantum Information and Quantum Physics, Chinese Academy of Sciences.

\section {Additional Information} Supplementary Information is available for this paper. Correspondence and requests for materials should be addressed to Hantao Sun or Xinfang Zhang(sunhantao@chinatelecom.cn or zhangxinfang@chinatelecom.cn).

\section{Data availability}
The data that support the findings of this study
are available at \url{https://qc.zdxlz.com/expExample/patchExpComplete?lang=zh}


\subsection*{\textit{\textbf{}{Tianyan}} Quantum Group}

Hantao Sun$^{1,2*}$, Xinfang Zhang$^{1,2*}$, Zhen Wang$^{1,2}$, Zilong Zha$^{1,2}$, Jie Chen$^{1,2}$, Qiankun Wang$^{1,2}$, Jiasheng Hu$^{1,2}$, Hanyi Wang$^{1,2}$, Jianjian Gao$^{1,2}$, Junhe Wang$^{1,2}$, Bihao Guo$^{1,2}$, Chunwang Liu$^{1,2}$, Yang Li$^{1,2}$, Yu Fan$^{1,2}$, Jian Ding$^{1,2}$, Jiping Fu$^{1,2}$, Yuanhang Hu$^{1,2}$, Rui Yang$^{1,2}$, Weizhi Tao$^{1,2}$, Zhehui Wang$^3$, Huijie Guan$^3$, ShuXin Xie$^3$, Liang Zhou$^3$, Jiafei Wang$^3$, LiangChao Sun$^3$, Zhong Zhao$^3$, Chang Wang$^3$, WenHao Chu$^3$, Shengbin Wang$^{1,2}$, Tao Yuan$^{1,2}$, Xixi Zhang$^{1,2}$,  Lei Miao$^{1,2}$, Wei Liu$^{1,2}$, Wenya Huang$^{1,2}$, Mengtian Zhou$^{1,2}$, Hanze Guo$^{1,2}$, He Huang$^{1,2}$, Yang Jin$^{1,2}$, Yin Kan$^{1,2}$, Zhaowei Li$^{1,2}$, Zirun Luo$^{1,2}$, Yuan Liu$^{1,2}$, Lijun Sun$^{1,2}$, Yue Tian$^{1,2}$, Mian Tao$^{1,2}$, Jiawang Wan$^{1,2}$, Rui Xu$^{1,2}$, Jiahao Xie$^{1,2}$, Shulin Yu$^{1,2}$, Yuting Zhang$^{1,2}$, Shuyue Zheng$^{1,2}$, Canxia Zhou$^{1,2}$, Tianqi Zong$^{1,2}$


$^1$ China Telecom Quantum Information Technology Group Co., Ltd., Hefei 230031, China

$^2$ Quantum Research Institute of China Telecom, Hefei 230031, China

$^3$ QuantumCTek Co., Ltd., Hefei 230026, China\\

$^*$ Corresponding authors (Hantao Sun: sunhantao@chinatelecom.cn, Xinfang Zhang: zhangxinfang@chinatelecom.cn)

\bibliography{main.bib}    

\begin{thebibliography}{10}

\bibitem{greenwade93}
George~D. Greenwade.
\newblock The {C}omprehensive {T}ex {A}rchive {N}etwork ({CTAN}).
\newblock {\em TUGBoat}, 14(3):342--351, 1993.

\bibitem{preskill2018}
John Preskill.
\newblock Quantum computing in the nisq era and beyond.
\newblock {\em Quantum}, 2:79, 2018.

\bibitem{arute2019quantum}
Frank Arute, Kunal Arya, Ryan Babbush, et~al.
\newblock Quantum supremacy using a programmable superconducting processor.
\newblock {\em Nature}, 574(7779):505--510, 2019.

\bibitem{wu2021strong}
Yulin Wu, Wan-Su Bao, Sirui Cao, et~al.
\newblock Strong quantum computational advantage using a superconducting
  quantum processor.
\newblock {\em Physical Review Letters}, 127(18):180501, 2021.

\bibitem{zhu2022quantum}
Qingling Zhu, Sirui Cao, Fusheng Chen, et~al.
\newblock Quantum computational advantage via 60-qubit 24-cycle random circuit
  sampling.
\newblock {\em Science Bulletin}, 67(3):240--245, 2022.

\bibitem{morvan2024phase}
Alexis Morvan, B~Villalonga, X~Mi, et~al.
\newblock Phase transitions in random circuit sampling.
\newblock {\em Nature}, 634(8033):328--333, 2024.

\bibitem{zhong2020quantum}
Han-Sen Zhong, Hui Wang, Yu-Hao Deng, et~al.
\newblock Quantum computational advantage using photons.
\newblock {\em Science}, 370(6523):1460--1463, 2020.

\bibitem{deng2023gaussian}
Yu-Hao Deng, Yi-Chao Gu, Hua-Liang Liu, et~al.
\newblock Gaussian boson sampling with pseudo-photon-number-resolving detectors
  and quantum computational advantage.
\newblock {\em Physical Review Letters}, 131(15):150601, 2023.

\bibitem{zhong2021phase}
Han-Sen Zhong, Yu-Hao Deng, Jian Qin, et~al.
\newblock Phase-programmable gaussian boson sampling using stimulated squeezed
  light.
\newblock {\em Physical Review Letters}, 127(18):180502, 2021.

\bibitem{madsen2022quantum}
Lars~S Madsen, Fabian Laudenbach, Mohsen~Falamarzi Askarani, et~al.
\newblock Quantum computational advantage with a programmable photonic
  processor.
\newblock {\em Nature}, 606(7912):75--81, 2022.

\bibitem{google2025}
Observation of constructive interference at the edge of quantum ergodicity.
\newblock {\em Nature}, 646(8086):825--830, 2025.

\bibitem{gao2025establishing}
Dongxin Gao, Daojin Fan, Chen Zha, et~al.
\newblock Establishing a new benchmark in quantum computational advantage with
  105-qubit zuchongzhi 3.0 processor.
\newblock {\em Physical Review Letters}, 134(9):090601, 2025.

\bibitem{17huang2024design}
He-Liang Huang, Youwei Zhao, and Chu Guo.
\newblock How to design a classically difficult random quantum circuit for
  quantum computational advantage experiments.
\newblock {\em Intelligent Computing}, 3:0079, 2024.

\bibitem{markov2008simulating}
Igor~L Markov and Yaoyun Shi.
\newblock Simulating quantum computation by contracting tensor networks.
\newblock {\em SIAM Journal on Computing}, 38(3):963--981, 2008.

\bibitem{guo2019general}
Chu Guo, Yong Liu, Min Xiong, et~al.
\newblock General-purpose quantum circuit simulator with projected
  entangled-pair states and the quantum supremacy frontier.
\newblock {\em Physical Review Letters}, 123(19):190501, 2019.

\bibitem{20villalonga2019}
Benjamin Villalonga, Sergio Boixo, Bron Nelson, et~al.
\newblock A flexible high-performance simulator for verifying and benchmarking
  quantum circuits implemented on real hardware.
\newblock {\em npj Quantum Information}, 5(1):86, 2019.

\bibitem{villalonga2020}
Benjamin Villalonga, Dmitry Lyakh, Sergio Boixo, et~al.
\newblock Establishing the quantum supremacy frontier with a 281 pflop/s
  simulation.
\newblock {\em Quantum Science and Technology}, 5(3):034003, 2020.

\bibitem{huang2020classical}
Cupjin Huang, Fang Zhang, Michael Newman, et~al.
\newblock Classical simulation of quantum supremacy circuits.
\newblock {\em arXiv preprint arXiv:2005.06787}, 2020.

\bibitem{pan2022simulation}
Feng Pan and Pan Zhang.
\newblock Simulation of quantum circuits using the big-batch tensor network
  method.
\newblock {\em Physical Review Letters}, 128(3):030501, 2022.

\bibitem{guo2021verifying}
Chu Guo, Youwei Zhao, and He-Liang Huang.
\newblock Verifying random quantum circuits with arbitrary geometry using
  tensor network states algorithm.
\newblock {\em Physical Review Letters}, 126(7):070502, 2021.

\bibitem{pan2022solving}
Feng Pan, Keyang Chen, and Pan Zhang.
\newblock Solving the sampling problem of the sycamore quantum circuits.
\newblock {\em Physical Review Letters}, 129(9):090502, 2022.

\bibitem{fu2024surpassing}
Rong Fu, Zhongling Su, Han-Sen Zhong, et~al.
\newblock Surpassing sycamore: Achieving energetic superiority through
  system-level circuit simulation.
\newblock In {\em SC24: International Conference for High Performance
  Computing, Networking, Storage and Analysis}, pages 1--20. IEEE, 2024.

\bibitem{zhao2025leapfrogging}
Xian-He Zhao, Zhong, Han-Sen, Feng Pan, et~al.
\newblock Leapfrogging sycamore: harnessing 1432 gpus for 7$\times$ faster
  quantum random circuit sampling.
\newblock {\em National Science Review}, 12(3):nwae317, 2025.

\bibitem{guan2024single}
Huijie Guan, Fei Zhou, Francisco Albarr{\'a}n-Arriagada, et~al.
\newblock Single-layer digitized-counterdiabatic quantum optimization for
  p-spin models.
\newblock {\em Quantum Science and Technology}, 10(1):015006, 2024.

\bibitem{cqlib_guide}
$Tianyan$~Quantum Group and ecosystem partners.
\newblock Cqlib software development kit.
\newblock {\em https://gitee.com/cq-lib/cqlib}, 2025.

\end{thebibliography}
\bibliographystyle{unsrt}

\appendix 

\section{System Calibration}\label{sec:AppendixA}

\subsection{$Tianyan$-287 quantum system}\label{sec:AppendixA.1}
\renewcommand{\thefigure}{S\arabic{figure}}
\setcounter{figure}{0}

\begin{figure*}[t!]
\centering
\includegraphics[width=0.7
\linewidth]{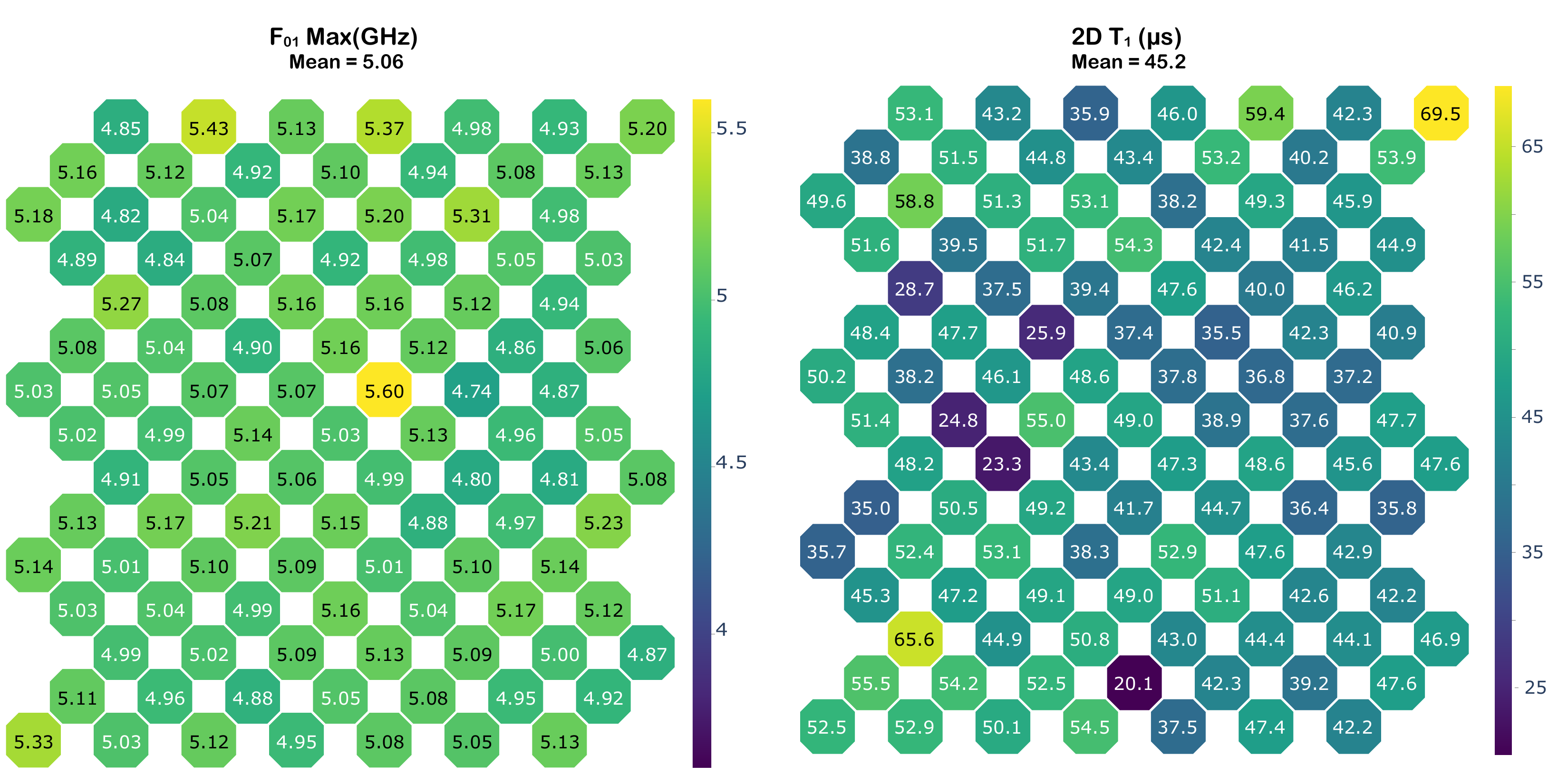}
\caption{\label{fig:104} Typical parameters distribution of the 104 working qubits of the $Tianyan$-287 system.}
\end{figure*}

\hspace{1.5em} The $Tianyan$-287 system features a $Zuchongzhi$-3.0-like quantum processor, comprising 105 qubits and 182 couplers. As shown in Fig.~\ref{fig:104}, the typical parameter distribution indicates that one qubit is non-functional, while two others exhibit suppressed T$_1$ relaxation times due to two-level systems. Over a 500 MHz frequency tunable range, two-dimensional (2D) T$_1$ measurements of  the remaining 104 qubits yield an average value of 45.2 $\mu$s.

\subsection{Calibration and Optimization for RCS}\label{sec:AppendixA.2}

\begin{figure*}[t!]
\centering
\includegraphics[width=0.95
\linewidth]{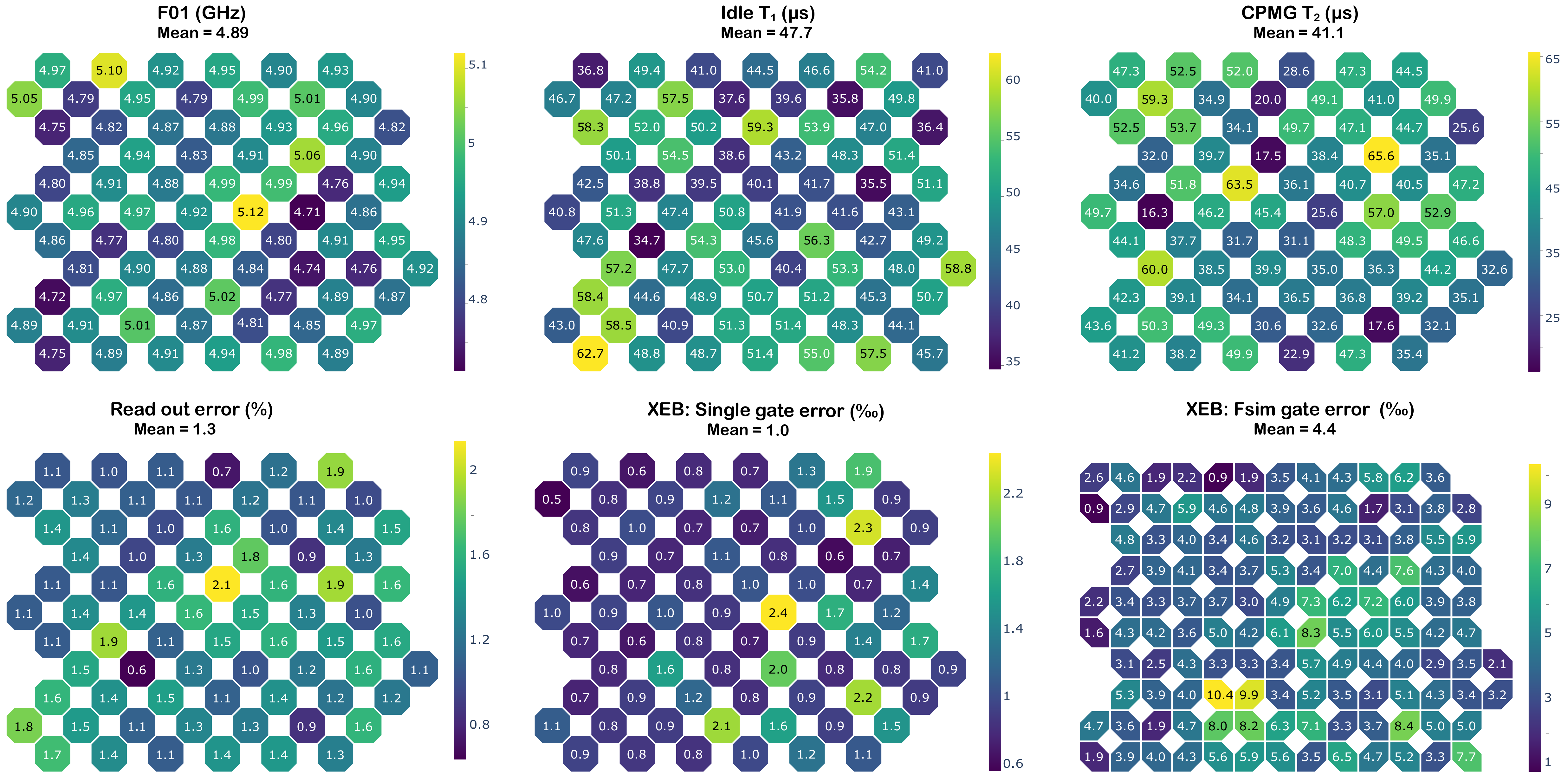}
\caption{\label{fig:param} Typical parameter distribution of the 74-qubit subset for RCS experiments.}
\end{figure*}

For the 74-qubit RCS task, the unused qubits were biased to the minimum operating frequency, and all inter-qubit couplings were deactivated to suppress unwanted interactions. To optimize the idle frequency configuration, a dynamic tuning strategy was determined by comprehensively considering factors such as coherence, two-level systems, and frequency collisions.
Experimental monitoring revealed that fluctuations in T$_1$ performance primarily originated from TLS effects. By performing real-time measurements of T$_1$ values near the idle frequencies and subsequent re-tuning, stable coherence times were ultimately achieved for the 74 qubits according to the frequency distribution of F01 (Fig.-\ref{fig:param}). This resulted in an average T$_1$ time of47.7 $\mu$s and a T$_2^{CPMG}$ time of 44.1 $\mu$s.

For single-qubit gate calibration, an optimal control theory-based scheme was utilized \cite{wu2021strong}. Under a fixed gate time of 26 ns, the drive frequency, amplitude, and DRAG coefficient were jointly optimized to minimize control errors. To mitigate crosstalk in large-scale systems, an anti-phase compensation cancellation technique was implemented. Verified through fully-parallel cross-entropy benchmarking (XEB), the single-qubit gates achieved an average fidelity of 99.9\% (Pauli error  = 1$\text{\textperthousand}$). In the readout stage, the X12-gate-driven 0-2 readout method \cite{morvan2024phase} was applied to effectively reduce decoherence errors during measurement. Furthermore, correlated readout errors were controlled below 2\% using a synchronous state preparation and measurement protocol.

For the implementation of two-qubit iSWAP-like gates \cite{gao2025establishing}, we first determined the optimal swap frequency via parameter scanning while fix the gate length at 40 ns, taking into account both coherence times and frequency collision thresholds. To address pulse distortion, a fine calibration method based on odd-numbered gate sequences was adopted to precisely adjust the coupling strength $g$ and detuning frequency. Given the characteristic of the four alternately executed iSWAP-like gate patterns (labeled A, B, C, and D), where only a subset of qubits is activated in each pattern, a dual compensation mechanism was established. By utilizing dynamic coupling-off technology, we successfully reduced the unintended residual coupling probability of adjacent couplers; additionally, idle gate benchmarking and calibration were introduced, significantly enhancing the stability of gate operations. Verification via XEB \cite{arute2019quantum} confirmed these improvements, with measurement results for the 74-qubit system across all patterns shown in Fig.-\ref{fig:param}.

\subsection{Cqlib}\label{sec:AppendixA.3}

Cqlib is an open-source quantum computing software development kit (SDK) developed by $Tianyan$ Quantum Group and ecosystem partners \cite{cqlib_guide}. Built upon the Quantum Computing Instruction Set (QCIS) as its core framework, it aims to provide convenient end-to-end technical support for quantum computing experiment design and application development, thereby lowering the barriers to the adoption and implementation of quantum computing technologies.

Its core capabilities cover key quantum computing needs: first, end-to-end quantum circuit processing, which can complete the full workflow from circuit construction, compilation and optimization to result output, meeting the experimental design requirements of different scenarios; second, high-efficiency simulation and visualization, supporting the simulation and verification of quantum circuits in classical environments, while providing visual displays of circuit structures and experimental results to help users intuitively understand quantum computing processes; third, cross-platform and cross-ecosystem compatibility, natively supporting hardware access to two major quantum computing cloud platforms (Tianyan and QuantumCTek), and enabling interoperability with mainstream quantum toolkits such as Qiskit, Cirq, and PennyLane through adapters, allowing users to directly execute circuits constructed with third-party toolkits on target hardware.

To further enhance the quantum computing ecosystem, the team is currently advancing the development of quantum machine learning libraries and quantum algorithm libraries. Through continuous iteration, Cqlib is committed to becoming a foundational software that bridges quantum hardware and upper-level applications, providing unified support for quantum computing practices in both academic and industrial communities.

\end{document}